\documentclass[a4paper,fleqn]{cas-dc}

\usepackage[numbers,sort&compress]{natbib}

\usepackage[export]{adjustbox}
\usepackage{booktabs}

\usepackage{array}
\newcolumntype{C}[1]{>{\centering\arraybackslash}p{#1}}

\def\tsc#1{\csdef{#1}{\textsc{\lowercase{#1}}\xspace}}
\tsc{WGM}
\tsc{QE}

\usepackage{listings}
\definecolor{backcolor}{rgb}{0.99,0.98,0.98}
\definecolor{string-color}{rgb}{0.3333, 0.5254, 0.345}
\definecolor{darkgrey}{rgb}{0.0627, 0.07, 0.082}
\definecolor{darkred}{rgb}{0.3, 0.05, 0.05}
\definecolor{codeblue}{rgb}{0.2,0.35,0.75}
\definecolor{codepurple}{rgb}{0.38,0.1,0.52}
\definecolor{codegray}{rgb}{0.5,0.5,0.5}
\definecolor{codegreen}{rgb}{0.05,0.3,0.05}
\definecolor{codered}{rgb}{0.6,0.2,0.1}
\definecolor{backgroundColour}{rgb}{0.99,0.99,0.98}
\definecolor{framegray}{rgb}{0.8,0.8,0.8}
\lstdefinestyle{myStyle}{
	language = C++,
	basicstyle = {\ttfamily \small \color{darkgrey}},
	backgroundcolor = {\color{backcolor}},
	commentstyle=\color{codegreen},
	stringstyle = {\color{string-color}},
	keywordstyle = {\color{codeblue}},
	keywordstyle = [2]{\color{codepurple}},
	keywordstyle = [3]{\color{codered}},
	keywordstyle = [4]{\color{codegray}},
	keywordstyle = [5]{\color{codegreen}},
	otherkeywords = {<, >, :, ::, DiFfRG, constexpr, uint, size_t, &, get, vector, array, Tensor, Scalar, FunctionND},
	morekeywords = [2]{AbstractModel, FEFunctionDescriptor, VariableDescriptor, ExtractorDescriptor, ComponentDescriptor
		TimeStepperSUNDIALS_IDA, UMFPack, Point, real, AD, NoJacobians, FE_AD,LLFFlux,FlowBoundaries
	},
	morekeywords = [3]{DiFfRG, CG, DG, dDG, LDG, def, Variables, dealii, std, autodiff},
	morekeywords = [4]{<, >, :, ::, ;, &},
	morekeywords = [5]{},
	breakatwhitespace=false,
	breaklines=true,
	captionpos=b,
	keepspaces=true,
	numbers=left,
	numbersep=5pt,
	numberstyle=\scriptsize\color{darkred},
	showspaces=false,
	showstringspaces=false,
	showtabs=false,
	tabsize=2,
	frame=single,
	rulecolor=\color{framegray},
	framerule=0.4pt,
	framesep=4pt,
	frameround=tttt
}
\lstdefinestyle{genStyle}{
	basicstyle = {\ttfamily \small \color{darkgrey}},
	backgroundcolor = {\color{backcolor}},
	commentstyle=\color{codegreen},
	stringstyle = {\color{string-color}},
	keywordstyle = {\color{codeblue}},
	breakatwhitespace=false,
	breaklines=true,
	captionpos=b,
	keepspaces=true,
	numbers=left,
	numbersep=5pt,
	numberstyle=\scriptsize\color{darkred},
	frame=single,
	rulecolor=\color{framegray},
	framerule=0.4pt,
	framesep=4pt,
	frameround=tttt,
	showspaces=false,
	showstringspaces=false,
	showtabs=false,
	tabsize=2
}
\lstdefinelanguage{CMake}{%
	morekeywords={if, else, endif, project, cmake_minimum_required, set, find_package, add_executable, include, target_link_libraries},%
	sensitive=false,%
	morecomment=[l]{\#},%
	morecomment=[s]{/*}{*/},%
	morestring=[b]",%
	otherkeywords={add_flows, setup_application},%
	keywordstyle = [2]{\color{codepurple}},
	morekeywords = [2]{REQUIRED, HINTS, VERSION, SYSTEM, LANGUAGES, PRIVATE},
}
\lstdefinelanguage{Bash}{%
	morekeywords={if, else, fi, mkdir, cd, cmake, bash, git},%
	sensitive=false,%
	morecomment=[l]{\#},%
	morecomment=[s]{/*}{*/},%
	morestring=[b]",%
	keywordstyle = {\color{codeblue}},
	keywordstyle = [2]{\color{codepurple}},
	morekeywords = [2]{$, /, ..},
}

\usepackage[nameinlink]{cleveref}

\usepackage{float}
\floatstyle{plain}
\newfloat{listing}{tbp}{lol}
\floatname{listing}{Listing}
\crefname{listing}{listing}{listings}
\Crefname{listing}{Listing}{Listings}
\usepackage{subcaption}
\DeclareCaptionSubType{listing}
\fnbelowfloat

\ExplSyntaxOn
\newcommand{\capfitwidth}{\dim_set:Nn \l_fig_width_dim { \linewidth }}
\ExplSyntaxOff

\newlength{\emailindent}
\newcommand{\caslinkmail}[1]{\href{mailto:#1}{\ttfamily\detokenize{#1}}}
\newcommand{\caslinkorcid}[1]{\href{https://orcid.org/#1}{\ttfamily\detokenize{#1}}}
\ExplSyntaxOn
\RenewDocumentCommand \emailauthor { m m }
   {
     \int_gincr:N \g_ead_int
     \seq_gput_right:Nn \g_stm_ead_seq
       {
         \caslinkmail { #1 }
         \parsename { #2 }
         \space(\eadauthor)%
       }
   }
\RenewDocumentCommand \printemails { }
{
  \group_begin:
  \int_compare:nNnTF { \int_use:N \g_ead_int } > { 0 }
  {
    \tex_let:D \thefootnote \relax \footnotetext
    {
      \raggedright
      \bool_if:NTF \g_stm_nologo_bool
      {
        \int_compare:nTF { \g_ead_int = 1 }
        { \settowidth \emailindent { \textit{Email~address:~} }
          \textit{Email~address:\c_space_token} }
        { \settowidth \emailindent { \textit{Email~addresses:~} }
          \textit{Email~addresses:\c_space_token} }
      }
      {
        \settowidth \emailindent
          { \includegraphics[height=8pt]{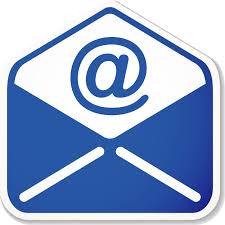}~ }
        \includegraphics[height=8pt]{thumbnails/cas-email.jpeg}\c_space_token
      }
      \dim_compare:nNnT { \footnotemargin } > { \c_zero_dim }
        { \addtolength \emailindent { \footnotemargin } }
      \hangindent \emailindent \hangafter 1
      \seq_use:Nn \g_stm_ead_seq { ;\newline }
    }
  }
  {  }
  \group_end:
}
\newlength{\orcidindent}
\RenewDocumentCommand \orcidauthor { m m }
   {
     \seq_gput_right:Nn \g_stm_orcid_seq
       {
         \caslinkorcid { #1 }
         \parsename { #2 }
         \space(\eadauthor)%
       }
   }
\RenewDocumentCommand \printorcid { }
{
  \group_begin:
  \seq_if_empty:NF \g_stm_orcid_seq
  {
    \tex_let:D \thefootnote \relax \footnotetext
    {
      \raggedright
      \settowidth \orcidindent { \textsc{orcid}(s):~ }
      \dim_compare:nNnT { \footnotemargin } > { \c_zero_dim }
        { \addtolength \orcidindent { \footnotemargin } }
      \textsc{orcid}(s):\c_space_token
      \hangindent \orcidindent \hangafter 1
      \seq_use:Nn \g_stm_orcid_seq { ;\newline }
    }
  }
  \group_end:
}
\ExplSyntaxOff

\newif\iflogoonlyinfobox
\logoonlyinfoboxtrue

\usepackage{zref-savepos}
\newlength{\abstractheight}
\newlength{\abstractextra}
\makeatletter
\let\cas@abstractname\abstractname
\renewcommand{\abstractname}{\zsavepos{casabsstart}\cas@abstractname}
\newcommand{\measureabstractheight}{%
  \setlength{\abstractheight}{\z@}%
  \zref@ifrefundefined{casabsstart}{}{%
    \zref@ifrefundefined{casabsend}{}{%
      \setlength{\abstractheight}{%
        \dimexpr\zposy{casabsstart}sp-\zposy{casabsend}sp+\abstractextra\relax}%
    }%
  }%
}
\newcommand{\logoboxspec}{\ifdim\abstractheight>\z@ to \abstractheight\fi}
\makeatother

\iflogoonlyinfobox
\ExplSyntaxOn
\RenewDocumentEnvironment { keywords } { O{} }
  {
    \measureabstractheight
    \tex_global:D \tex_setbox:D \g_stm_key_box = \vtop \bgroup
    \hsize=.3 \textwidth
    \parindent 0pt
    \vbox \logoboxspec \bgroup
    \vfil
  }
  { \vfil \egroup \egroup }
\ExplSyntaxOff
\fi

\newcommand{\TempLat}{\texttt{TempLat}\xspace}
\newcommand{\CosmoLattice}{\texttt{CosmoLattice}\xspace}

\newcommand{\ParaFaFT}{\texttt{ParaFaFT}\xspace}
\newcommand{\Kokkos}{\texttt{Kokkos}\xspace}
\newcommand{\HILA}{\texttt{HILA}\xspace}
\newcommand{\cpp}[1]{\lstinline[language=C++,style=myStyle,frame=none]|#1|}
\newcommand{\cmake}[1]{\lstinline[language=CMake,frame=none]|#1|}
\newcommand{\bash}[1]{\lstinline[language=Bash,frame=none]|#1|}

\newsavebox{\widestdecl}
\newsavebox{\widestobj}

\graphicspath{{./figures/}{../bench-hila-templat/analysis/figs/}}

\newlength{\pairfigwidth}
\newlength{\pairfigwide}
\newlength{\pairfignarrow}
\begin{document}
\let\WriteBookmarks\relax
\def\floatpagepagefraction{1}
\def\textpagefraction{.001}

\shorttitle{TempLat: a versatile C++ engine for lattice field theories}

\shortauthors{A. Florio and F. R. Sattler}

\title [mode = title]{TempLat: a versatile C++ engine for lattice field theories \\ {\it \large Powering CosmoLattice on GPUs}}  




\author[1]{Adrien Florio}[orcid=0000-0002-7276-4515]



\ead{adrien.florio@uni-bielefeld.de}



\affiliation[1]{organization={Fakult{\"a}t f{\"u}r Physik},
            addressline={Universit{\"a}t Bielefeld}, 
            city={Bielefeld},
            postcode={D-33615}, 
            country={Germany}}

\author[1]{Franz R. Sattler}[orcid=0000-0003-1744-9456]


\ead{fsattler@physik.uni-bielefeld.de}








\begin{abstract}
We present \TempLat, a C++ framework for lattice field theory simulations in arbitrary dimensions. Its symbolic language, built on expression templates, translates mathematical expressions almost verbatim into C++ while compiling to highly optimized kernels, in contexts ranging from classical-statistical to Monte Carlo simulations. 
\TempLat is performance-portable: building on \Kokkos, it supports large CPU clusters, as well as NVIDIA and AMD GPUs. Its abstraction of hardware into a device concept makes \TempLat extensible to future architectures.
We demonstrate excellent strong and weak scaling on both CPUs and GPUs. We also present \ParaFaFT, a standalone parallel discrete Fourier transform library supporting arbitrary dimensions on all of the above hardware, and benchmark its scaling. Both are open source and available on GitHub, and power a new release of \CosmoLattice, a widely used early-universe simulation library now available on GPUs. \zsavepos{casabsend}%
\end{abstract}



\begin{keywords}
\iflogoonlyinfobox
\noindent\href{https://cosmolattice.github.io/templat/}{\includegraphics[width=\hsize]{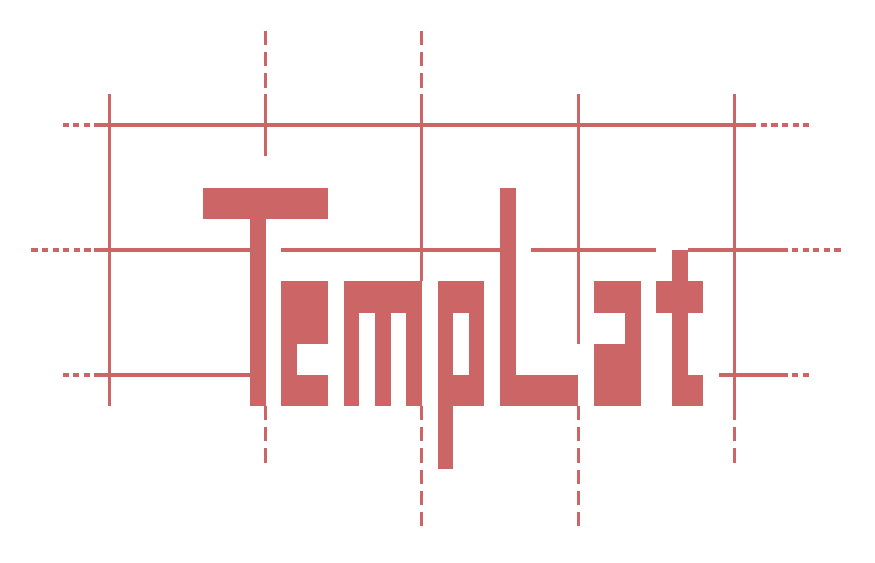}}
\else
Lattice field theory \sep Expression templates \sep Performance portability \sep GPU computing
\par\vspace{0.8em}
\noindent\href{https://cosmolattice.github.io/templat/}{\includegraphics[width=.966\hsize]{templat_logo}}
\fi
\end{keywords}

\maketitle


\section{Introduction}
\label{sec:introduction}

Field theories are used to describe a large variety of phenomena, ranging from hydrodynamics to the Standard Model of particle physics. In most cases, purely analytical tools can probe only limited regimes of the full theories, and numerical methods are indispensable.
One such case is the study of early-universe dynamics, when the fate of the universe is dictated by classical field dynamics. Another is the study of critical phenomena and their impact on dynamics. A third is that of strongly coupled, non-perturbative theories like quantum chromodynamics, the theory of the strong force in the Standard~Model.

A powerful technique common to all of these is the lattice discretization of space(-time). In quantum chromodynamics, Monte Carlo simulations on space-time lattices are employed \cite{Egri:2006zm, Clark:2009wm, OpenQCD, Sexty:2013ica, Boyle:2015tjk, Altenkort:2021cvg, HotQCD:2023ghu}.
Lattice discretizations are likewise used in hydrodynamics \cite{Incompact3D} and magneto-hydrodynamics \cite{PencilCode:2020eyn, Stone2020}, Landau--Lifshitz micromagnetics \cite{bruckner2023magnumnppytorchbased, mumax3}, and classical statistical simulations of cold-atom systems \cite{PineiroOrioli:2015cpb}.
All these simulations need massively parallelized representations of discrete fields. Cross-fertilization calls for efficient and versatile all-around software that goes beyond specialized uses.

\TempLat is one such library. It began as, and still is, the C++ engine behind \CosmoLattice \cite{Figueroa:2020rrl,Baeza-Ballesteros:2025tme,Figueroa:2021yhd,Figueroa:2023xmq}, a widely used early-universe simulation stack (see \url{https://cosmolattice.net} for a live list of publications that use it). The evolution of the HPC landscape called for a redesign supporting GPU architectures. This led \TempLat to mature into a standalone library, whose first release we present here and which is available on GitHub\footnote{\url{https://github.com/cosmolattice/templat}}. Built on expression templates, it fully decouples abstract mathematical expressions from their mapping onto hardware.
Its core engine builds on \Kokkos \cite{KokkosCore2014, KokkosEcosystem2021, KokkosCore2022}, which seamlessly dispatches computations to different CPU and GPU architectures and from whose active development \TempLat continuously benefits. Decoupling \TempLat from \CosmoLattice achieves several aims. First, it separates faster-paced improvements to the core engine from the slower-paced release of new cosmological models. 
Second, it provides the community with a complementary high-performance code
alongside \HILA \cite{HILA}, a platform-agnostic C++ engine that has run
$16\,384^3$ scalar field simulations on the AMD LUMI
cluster~\cite{Correia:2024cpk}, with $24\,576^3$ runs
underway~\cite{Rummukainen:private}.
Having two independent codes makes cross-validation possible. Finally, the success of \CosmoLattice demonstrates that \TempLat is genuinely accessible: it has brought lattice simulations to a large community of non-specialists. The standalone \TempLat is designed to carry that accessibility beyond cosmology.

To explain how \TempLat achieves this, we introduce its interface in \cref{sec:interface} and its architecture in \cref{sec:how}. We demonstrate its performance in \cref{sec:bench} and conclude with an outlook in \cref{sec:conclusions}.

\section{\TempLat: Interface}
\label{sec:interface}
A particular strength of \TempLat is its expression template system, which allows translating mathematical expressions almost verbatim into C++ code.
To illustrate this, consider the Langevin equation for a scalar $\phi^4$ theory in $d=3$. The Langevin equation is given as
\begin{align}\label{eq:Langevin}
	\partial_t \phi = - \frac{\delta S}{\delta \phi} + \eta = \Delta\phi - \frac{\lambda}{2}\phi^3 + \eta\,,
\end{align}
with $\eta$ being Gaussian fluctuations with $\langle\eta(\boldsymbol{x},t)\eta(\boldsymbol{x}',t')\rangle = \delta(\boldsymbol{x}-\boldsymbol{x}')\delta(t-t')$.
Using \TempLat, this expression can be translated to C++ as,
\begin{lstlisting}[language=C++, style=myStyle, numbers=none]
auto expr = LatLapl(phi) - lambda / 2 * pow<3>(phi) 
             + eta;
\end{lstlisting}
Here, \cpp{expr} does \textit{not} contain the three-dimensional distribution of $\partial_t \phi$; it is a very lightweight object that symbolically encodes the mathematical expression.
Specifically, it is an abstract expression tree whose leaves consist of the following:

\begin{table}[pos=H]
	\centering
	\setlength{\tabcolsep}{0pt}
	\lstset{basicstyle={\ttfamily\footnotesize\color{darkgrey}}}
	\begin{lrbox}{\widestobj}
		\lstinline[frame=none]$LatLapl$
	\end{lrbox}
	\begin{tabular}{@{} l @{\hspace{25pt}} >{\raggedright\arraybackslash}p{\dimexpr\columnwidth-\wd\widestobj-25pt\relax} @{}}
	\toprule
	\textbf{Object} & \textbf{Type and description} \\
	\midrule
	\lstinline[frame=none]$LatLapl$ & \lstinline[frame=none]$LatticeLaplacian<E>$, the lattice-discretized Laplacian, wrapping the expression \lstinline[frame=none]$E$ it is applied to. \\
	\addlinespace[2pt]
	\lstinline[frame=none]$phi$ & \lstinline[frame=none]$Field<T, NDim>$, a lattice field which holds memory. \\
	\addlinespace[2pt]
	\lstinline[frame=none]$lambda$ & \lstinline[frame=none]$double$, a constant numerical value. \\
	\addlinespace[2pt]
	\lstinline[frame=none]$eta$ & \lstinline[frame=none]$RandomGaussianFieldConfig<T, NDim>$, a field-valued random number generator. \\
	\bottomrule
	\end{tabular}
\end{table}

As already alluded to, the expression \cpp{expr} is only evaluated on assignment to actual memory, in practice through the assignment operator \cpp{=} of specific types, see below. 
This enables optimization to be performed by walking the expression tree, before any assignment is made. This results in expressions that are optimized at compile time. Consider, for example, multiplying \cpp{expr} by $1$ or $0$. These two values have special dedicated types: \cpp{OneType} and \cpp{ZeroType} respectively. A multiplication by \cpp{ZeroType} collapses the entire expression tree to exactly $0$, i.e. \cpp{ZeroType}:
\begin{lstlisting}[language=C++, style=myStyle, numbers=none]
// The compile-time type of expr0 will be ZeroType
auto expr0 = expr * ZeroType();
\end{lstlisting}
Similarly, a multiplication by \cpp{OneType} will simply return the expression itself. This way, \TempLat does not perform unnecessary computations, but prunes expression trees already at compile time.

In \TempLat, any lattice simulation needs an environment which provides lattice sizes, global memory management and memory layouts. Therefore, the first step to any \TempLat simulation is to create a \cpp{MemoryToolBox}, e.g.
\begin{lstlisting}[language=C++, style=myStyle, numbers=none]
// Create a 128^3 box with one ghost cell
constexpr size_t NDim = 3;
const size_t N = 128;
const size_t ghosts = 1;
auto toolBox = MemoryToolBox<NDim>::makeShared(N, ghosts);
\end{lstlisting}
Notice that \TempLat is dimension-agnostic --- by changing the compile-time variable \cpp{NDim}, one can run this example in any number of dimensions $d \geq 1$.

With a \cpp{toolBox} defined, we can now define one of the central elements of \TempLat: the \cpp{Field} class, which represents a scalar field that lives on the lattice
\begin{lstlisting}[language=C++, style=myStyle, numbers=none]
Field<double, NDim> phi("phi", toolBox);
\end{lstlisting}
It serves two purposes. First, it can participate in the algebra and be a leaf of any expression template. Second, its assignment operator \cpp{=} triggers the evaluation of expressions across the whole lattice. Both purposes are achieved simultaneously by the \cpp{Field} holding a shared reference to a \cpp{MemoryBlock}, which owns actual memory on the computational device.  As a result, copies of \cpp{Field} are cheap (as for most classes in \TempLat), as they essentially just represent shared views to actual memory data. This is indeed a key requirement to be able to take part in the algebra.

In more detail, the assignment operator \cpp{=} of a \cpp{Field} (or the complex, $U(1)$ or $SU(2)$ analogues, see below) operates by default in position space. However, the class exposes a Fourier-space view, to perform assignments in momentum space:
\begin{lstlisting}[language=C++, style=myStyle, numbers=none]
// Fourier-transforms to momentum space 
// and assigns 4 to the memory
phi.inFourierSpace() = 4;
// Fourier-transforms back, evaluates the
// expression, assigns the result to phi
phi = expr;
\end{lstlisting}
As a result, there is no need to invoke Fourier transformations manually; \TempLat automatically performs these when necessary.

\begin{sloppypar}
Another useful class we encountered in the example is \cpp{RandomGaussianFieldConfig<T, NDim>}, which represents Gaussian random fluctuations in configuration space (the corresponding Fourier-space class is \cpp{RandomGaussianField<T, NDim>}). More generally, \TempLat provides classes that represent field-valued random number generators.
It uses the \texttt{Random123} \cite{10.1145/2063384.2063405} library, specifically, the counter-based 2$\times$64-bit Philox algorithm. A great benefit of the statelessness of this generator is that the generation of random numbers can be done independently on every single lattice site, which leads to very fast generation times as compared to typical stateful RNGs. It also makes full reproducibility easy to achieve. 
\end{sloppypar}

Altogether, \cref{lst:langevin-example} shows the full example of evolving a system with $128^3$ points using the Langevin equation \eqref{eq:Langevin}
, where the noise is rescaled by $1/\sqrt{\mathrm{d}t}$ to get the correct $\mathrm{d}t\to 0$ limit.
\begin{listing}[b]
\begin{lstlisting}[language=C++, style=myStyle, numbers=none]
// Create a 128^3 box with one ghost cell
constexpr size_t NDim = 3;
const size_t N = 128;
const size_t ghosts = 1;
auto toolBox = MemoryToolBox<NDim>::makeShared(N, ghosts);

// TempLat supports float and double number types
using T = double;

// Field classes hold data
Field<T, NDim> phi("phi", toolBox);

// Create an object giving gaussian fluctuations
RandomGaussianFieldConfig<T, NDim> eta("seed", toolBox);

// Create an abstract expression that can be
// evaluated on the lattice; the noise is rescaled
// by 1/sqrt(dt) to obtain the discretized
// delta correlation in time
const T lambda = 0.1;
const T dt = 0.01;
auto expr = LatLapl(phi) - lambda / 2 * pow<3>(phi)
            + sqrt(1 / dt) * eta;

// Now perform 100 evolution steps
for(size_t step = 0; step < 100; ++step) {
  // This evaluates the expression and assigns it to the data held by phi
  phi = phi + dt * expr;
}
\end{lstlisting}
\caption{Langevin evolution of a scalar field, \cref{eq:Langevin}.}
\label{lst:langevin-example}
\end{listing}
\begin{table*}[t]
	\centering
	\setlength{\tabcolsep}{0pt}
	\lstset{basicstyle={\ttfamily\footnotesize\color{darkgrey}}}
	\begin{lrbox}{\widestdecl}
		\lstinline[frame=none]$auto st = ConstructSymTraceless(t11, t12, t13, t22, t23, t33);$
	\end{lrbox}
	\begin{tabular}{@{} l @{\hspace{12pt}} p{\dimexpr\textwidth-\wd\widestdecl-12pt\relax} @{}}
	\toprule
	Declaration & Description \\
	\midrule
	\multicolumn{2}{@{}l}{\emph{Field types}, all constructed from a name tag and a \texttt{MemoryToolBox}} \\
	\addlinespace[2pt]
	\lstinline[frame=none]$ComplexField<T, NDim> cField("cField", toolBox);$ & a complex scalar field \\
	\lstinline[frame=none]$U1Field<T, NDim> u1Field("u1Field", toolBox);$ & a single $U(1)$ link $U_i = e^{\mathrm{i} A_i}$ for fixed $i$; an alias of \texttt{ComplexField}, as $U(1)$ group elements are unit-modulus complex numbers \\
	\lstinline[frame=none]$SU2Field<T, NDim> su2Field("su2Field", toolBox);$ & a single $U^a_i$ for fixed $i$, with $a=1\dots4$ \\
	\lstinline[frame=none]$SU2LieAlgebraField<T, NDim> su2Alg("su2Alg", toolBox);$ & an $\mathfrak{su}(2)$ field, i.e.\ the three generators $a=1,2,3$ \\
	\lstinline[frame=none]$SU2Doublet<T, NDim> doublet("doublet", toolBox);$ & an $SU(2)$ (Higgs) doublet, i.e.\ four real components \\
	\lstinline[frame=none]$SymTracelessField<T, NDim> hij("hij", toolBox);$ & a real symmetric traceless $3\times3$ matrix field (five components) \\
	\addlinespace
	\midrule
	\multicolumn{2}{@{}l}{\emph{Containers}, bundling any of the above under a single tag-indexed object} \\
	\addlinespace[2pt]
	\lstinline[frame=none]$VectorField<FieldType> Us("Us", toolBox);$ & a vector-valued field, e.g.\ the link $U_\mu$ with $\mu = 1 \dots \texttt{NDim}$ \\
	\lstinline[frame=none]$FieldCollection<FieldType, N> coll("coll", toolBox);$ & \lstinline[frame=none]$N$ fields of arbitrary type \\
	\lstinline[frame=none]$VectorFieldCollection<FieldType, N> Ws("Ws", toolBox);$ & nests the two, e.g.\ for a collection of gauge fields \\
	\addlinespace
	\midrule
	\multicolumn{2}{@{}l}{\emph{Matrix expressions}, whose arguments are fields, expressions, or \texttt{ZeroType}/\texttt{OneType}} \\
	\addlinespace[2pt]
	\begin{tabular}[t]{@{}l@{}}
		\lstinline[frame=none]$auto mat = ConstructMatrix3x3(m11, m12, m13,$ \\
		\lstinline[frame=none]$                              m21, m22, m23,$ \\
		\lstinline[frame=none]$                              m31, m32, m33);$
	\end{tabular} & a general $3\times3$ matrix, given row by row \\
	\lstinline[frame=none]$auto sym = ConstructSym(s11, s12, s13, s22, s23, s33);$ & a symmetric $3\times3$ matrix, given by its upper triangle \\
	\lstinline[frame=none]$auto herm = ConstructHerm(h11, h12, h13, h22, h23, h33);$ & a hermitian $3\times3$ matrix, given by its upper triangle \\
	\lstinline[frame=none]$auto st = ConstructSymTraceless(t11, t12, t13, t22, t23, t33);$ & as \texttt{ConstructSym}, but with the trace subtracted upon evaluation \\
	\bottomrule
	\end{tabular}
	\caption{Field types, containers and matrix expressions provided by \TempLat, beyond the real scalar \texttt{Field}. The containers are the natural types for gauge links and multiplets; the matrix expressions are again lightweight expression objects rather than storage, so that entries set to \texttt{ZeroType} are pruned from the expression tree at compile time.}
	\label{tab:field-types}
\end{table*}
Besides scalar fields and their corresponding algebra, \TempLat provides the field types, index-carrying containers and matrix expressions collected in \cref{tab:field-types}. Having composite types allows us to define specialized algebra for these different types. For instance, \cpp{SU2Field}s have their own assignment operators \cpp{=} which act and assign $SU(2)$ matrices directly. This goes hand in hand with the definition of a separate expression-template $SU(2)$ algebra, which implements quaternion multiplication, the $SU(2)$ exponential maps and other relevant operations. All other composite types also come with their own algebra. This architecture renders \TempLat very modular and easy to extend by adding new composite fields together with their own algebra.

Last, we want to highlight \textit{shifts}, one of the most important symbolic operations on a lattice. They are used to access fields whose position is shifted by a constant vector
\begin{align}
	\phi(x + \hat\mu) = \texttt{shift(phi, mu)}
\end{align}
where \cpp{mu} is a direction tag, e.g. \cpp{Tag<1>\{\}} for the first spatial direction.
They are for instance crucial to define finite-difference based operators. To further exemplify their use, consider the gauge-invariant ``plaquette'' term made out of $SU(2)$ fields 
\begin{align}
	P_{\mu\nu}(x) =	U_\mu(x)\,U_\nu(x+\hat \mu)\, U_\mu^\dagger(x+\hat \nu)\,U_\nu^\dagger(x)\,.
\end{align}
It can be elegantly written using the shift operator. To represent the links, we use the \TempLat definition of an \cpp{NDim}-component vector of $SU(2)$ fields, \cpp{VectorField<SU2Field<T, NDim>>}, and simply get 
\begin{lstlisting}[language=C++, style=myStyle, numbers=none]
VectorField<SU2Field<T, NDim>> Us("Us", toolBox);
// ...
auto exprPlaquette = (Us(mu) * shift(Us(nu), mu)) * (dagger(shift(Us(mu), nu)) * dagger(Us(nu)));
\end{lstlisting}
%

\subsection{Using \TempLat in user code}

Writing a \TempLat application requires wiring \TempLat into the CMake project of the application.
%
\begin{lstlisting}[language=CMake, numbers=none]
cmake_minimum_required(VERSION 3.16)
project(my_project LANGUAGES CXX)

include(FetchContent)
FetchContent_Declare(TempLat
  GIT_REPOSITORY 
      https://github.com/cosmolattice/templat.git
  GIT_TAG 
      main)
FetchContent_MakeAvailable(TempLat)

# Add your executable and link it against TempLat
add_executable(my_executable main.cpp)
target_link_libraries(my_executable 
                          PRIVATE TempLat::TempLat)
\end{lstlisting}
%
This will automatically fetch \TempLat from the GitHub repository when the project is configured, using \cmake{FetchContent}. The only thing required to use \cmake{my_executable} with \TempLat is to link it against the \cmake{TempLat::TempLat} library. \Cref{lst:main-example} shows a minimal working example \bash{main.cpp} to build a \TempLat executable.

Once both \bash{CMakeLists.txt} and \bash{main.cpp} are in a given folder, building the simulation goes as follows:
\begin{lstlisting}[language=bash, numbers=none]
mkdir build
cd build
cmake ..
make -j$(nproc)
\end{lstlisting}
The above builds a program that can be run as \bash{./my_executable} and yields the following output:
\begin{lstlisting}[numbers=none]
<@\textcolor{codeblue}{(0.000s)}@>
Using KokkosFFT backend for FFTs.

<@\textcolor{codeblue}{(0.000s) MPI Rank 0 - memorytoolbox.h:209 -->}@>
Using the following domain decomposition:(1 1 1 ).
<@\textcolor{codeblue}{(0.002s) MPI Rank 0 - main.cpp:15 -->}@>
Hello, TempLat!  <phi> = 1.99973
\end{lstlisting}
As \cpp{RandomGaussianFieldConfig} is based on a pseudo-random number generator shipped with \TempLat, the output will not be platform-dependent.

Note that by default \TempLat will try to configure itself optimally for the given platform. Explicitly, this means that \TempLat will probe software to enable specific devices, in the following order:
\begin{enumerate}
	\item[(i)] First, a GPU backend is chosen, depending on whether the corresponding software (CUDA or HIP toolkits) is available, in the following order:
	\begin{enumerate}
		\item CUDA\footnote{The compiler support of the CUDA toolkit can be found in the online documentation, e.g. at \url{https://docs.nvidia.com/cuda/archive/12.8.0/cuda-installation-guide-linux/index.html\#host-compiler-support-policy}, for other versions of CUDA simply change the version string 12.8.0 in this URL to the pertinent CUDA version.}
		\item HIP
		\item If neither is available, no GPU backend is enabled.
	\end{enumerate}
	\item[(ii)] Then, a CPU backend is chosen, even if a GPU has been detected:
	\begin{enumerate}
		\item OpenMP
		\item POSIX Threads
		\item If none of the above is found, the serial implementation is enabled.
	\end{enumerate}
\end{enumerate}
The automatic detection is only performed if no backends are explicitly enabled. For example, configuring the above explicitly with CUDA and a sequential CPU backend is done as follows:
\begin{lstlisting}[language=bash, numbers=none]
cmake .. -DCUDA=ON -DNOTHREADING=ON
\end{lstlisting}
Note that MPI has to be enabled by hand, i.e.\ it is not automatically detected:
\begin{lstlisting}[language=bash, numbers=none]
cmake .. -DMPI=ON
\end{lstlisting}
\begin{listing}[b]
	\begin{lstlisting}[language=C++, style=myStyle, numbers=none]
#include "TempLat.h"
		
int main(int argc, char *argv[])
{
  using namespace TempLat;
  SessionGuard guard(argc, argv);
  auto toolBox = 
      MemoryToolBox<3>::makeShared(128, 1);
  Field<double, 3> phi("phi", toolBox);
  phi = 2 + RandomGaussianFieldConfig<double, 3>(
                  "seed", toolBox);
  const double result = average(phi);
  sayMPI << "Hello, TempLat!  <phi> = "
         << result << "\n";
  return 0;
}
\end{lstlisting}
	\caption{A minimal \TempLat application, \texttt{main.cpp}.}
	\label{lst:main-example}
\end{listing}
%

\subsection{Unit tests}

\TempLat has high test coverage for all aspects of its codebase, which both ensures correctness and guards against future regressions. To enable the \TempLat tests, one needs only to switch them on:
\begin{lstlisting}[language=bash, numbers=none]
cmake -DTEMPLAT_TEST=ON ..
make -j$(nproc)
\end{lstlisting}
The above will enable tests and build all of them. Running the tests can be done through ctest:
\begin{lstlisting}[language=bash, numbers=none]
ctest --output-on-failure  # run all tests
ctest -R fft               # run tests named *fft*
\end{lstlisting}
%

\section{\TempLat: Architecture}
\label{sec:how}

\subsection{Hybrid memory parallelization}
\label{sec:parallel}

\begin{figure*}[t]
	\centering
	\includegraphics[width=0.6\linewidth]{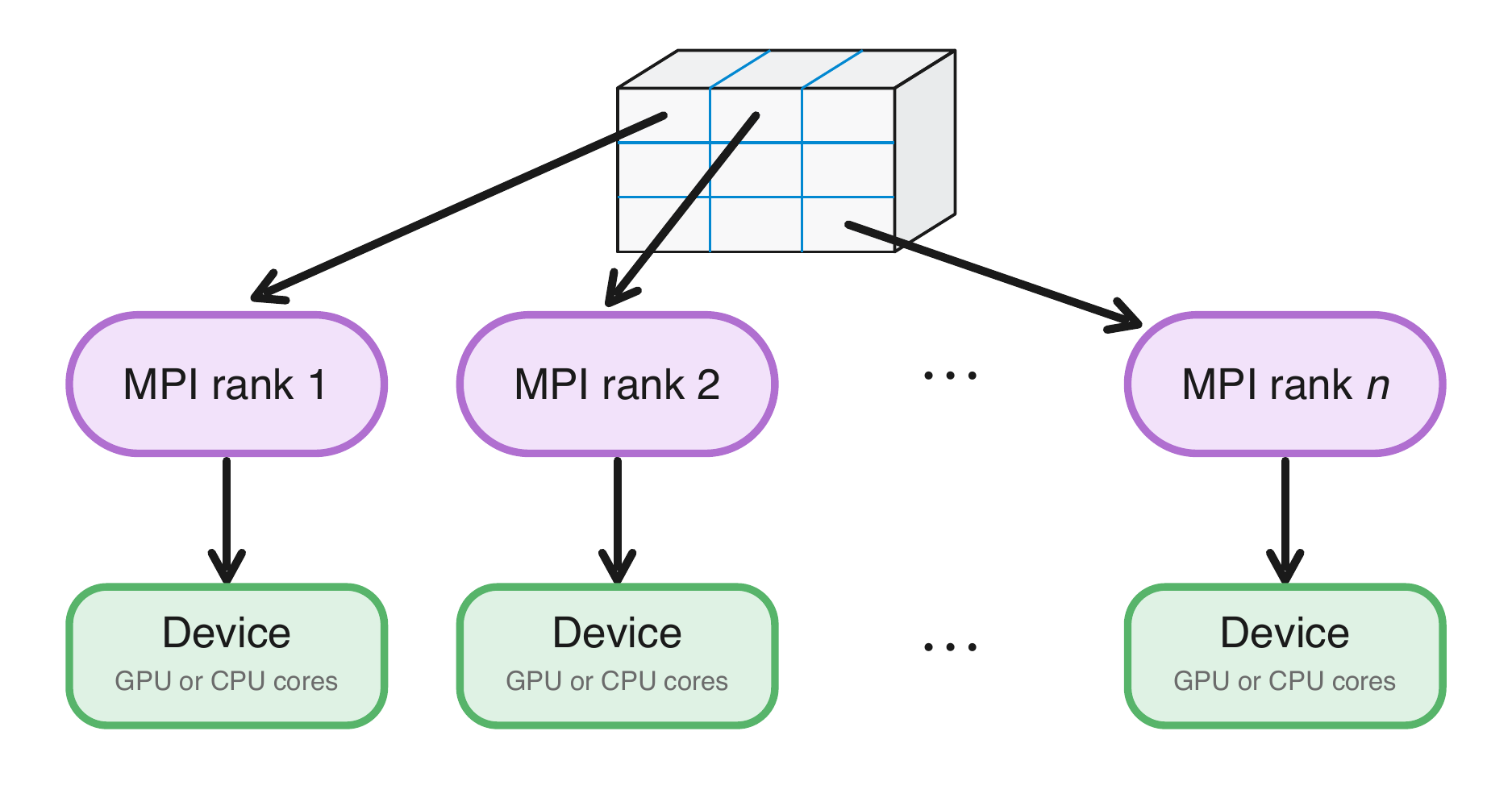}
	\caption{Hybrid parallelization structure}
	\label{fig:distributed_shared}
\end{figure*}

\TempLat implements a hybrid parallelization structure, which allows the simultaneous use of parallelization within a single computational device (e.g. a single CPU or GPU with many cores) and parallelization across machines connected through a network (distributed parallelization).

\textbf{Distributed} parallelization works across separate machines (nodes). This is implemented using the Message Passing Interface (MPI), which allows sending data packets across a network. As network communication in this case is expensive, the workload has to be split cleanly across the nodes, ensuring minimal communication overhead. In lattice simulations, the overall simulation domain can be cleanly split between nodes. The only necessary memory exchange is that of boundary cells, allowing the representation of interactions between nearest neighbors or even at larger distances. 
This mode of parallelization is necessary to scale up to large HPC clusters, in which hundreds or even thousands of nodes can work together on a single lattice simulation.
 
\textbf{Shared-memory} parallelization happens within a single machine, where all processes can access the same memory, without the need to copy or transfer data. In particular, this allows for very fine-grained parallelization, with every single lattice site processed in parallel. This is possible because most operations on the lattice are \textit{embarrassingly parallelizable}.
This fact opens the avenue of using massively parallel hardware such as GPUs for lattice simulations. \TempLat abstracts the hardware/software pair used for shared-memory parallelization as a \textit{device}. As of now, \TempLat supports the following devices:
 
\begin{table}[pos=H]
	\centering
	\begin{tabular}{@{} C{3cm} C{3cm} @{}}
		\toprule
		\textbf{Hardware} & \textbf{Software} \\
		\midrule
		AMD GPU & HIP \\
		NVIDIA GPU & CUDA \\
		\midrule
		\multirow{3}{*}{CPU} & OpenMP \\
		 & POSIX Threads \\
		 & Serial \\
		\bottomrule
	\end{tabular}
\end{table}
 
\noindent
To implement support for a wide range of devices, we use the \Kokkos performance-portability framework \cite{KokkosCore2014,KokkosEcosystem2021,KokkosCore2022}, which provides building blocks for portable code that can be compiled and run on any of the above~devices.

We link these two concepts and implement \textbf{hybrid parallelism} in \TempLat. A large lattice simulation splits over a set of $n$ MPI ranks, where $n$ is no larger than $N^{d-1}$ for a cubic lattice of extent $N$: the last direction is kept intact and the remaining $d-1$ are segmented (see \cref{sec:bench}). Each of these MPI ranks manages one device for computations --- either a GPU or a set of CPU cores --- which provides shared-memory parallelization locally on a given node, see also \cref{fig:distributed_shared}.

\TempLat also automatically probes the node topology to detect whether any subset of GPUs is directly connected via Peripheral Component Interconnect Express (PCIe). In this case, \TempLat lets neighboring GPUs exchange data directly over PCIe, yielding highly efficient intra-node communication. This efficiency is further increased when NVIDIA's NVLink or AMD's xGMI is available, which is likewise detected automatically.
\subsection{The host/device concept}
\label{sec:hostdevice}

\begin{listing*}[b]
\begin{lstlisting}[language=C++, style=myStyle, numbers=none]
	auto functor = DEVICE_CLASS_LAMBDA(const device::IdxArray<NDim> &idx) {
		device::apply([&](auto &&...args) {
			mView(args...) = DoEval::eval(g, args...);
		}, idx);
	};
	device::iteration::foreach("ConfigViewAssign", mLayout, functor);
\end{lstlisting}
\caption{A site-wise iteration over a given lattice layout \cpp{mLayout}.}
\label{lst:foreach}
\end{listing*}
\begin{listing*}[b]
\begin{lstlisting}[language=C++, style=myStyle, numbers=none]
	vType localResult{};
	auto functor = DEVICE_CLASS_LAMBDA(const device::IdxArray<NDim> &idx, vType &update) {
		device::apply([&](auto &&...args) {
			update = device::max(DoEval::eval(mT, args...), update);
		}, idx);
	};
	device::iteration::reduce("Maximum", mLayout, functor, device::iteration::Max<vType>(localResult));
\end{lstlisting}
\caption{A reduction over a given lattice layout \cpp{mLayout}, evaluating the maximum into \cpp{localResult}.}
\label{lst:reduce}
\end{listing*}

In addition to the device concept, we also use the concept of a \textit{host}.
The host always uses CPU hardware. Its purpose is to assemble the overall logic of the program, including management of resources, assembling expressions and dispatching these expressions to the device. 
The device then simply takes commands from the host and executes these on the hardware, modifying its managed memory. Depending on the hardware, the actual realization of these concepts is slightly different:
\begin{itemize}
	\item When using \textbf{GPU}-based devices, the separation between host (a CPU) and the device (the GPU) is both hardware- and software-based. Importantly, the device side owns all the memory associated with the simulation and executes host-issued commands on this memory.
	As memory copies between GPU and CPU memory are very slow and expensive, \TempLat never performs these on large memory blocks --- only very small results of computations are ever moved between the memories of host and device.
	\item In the case of a \textbf{CPU}-based device, the distinction is purely virtual as the hardware used for host and device is the same, only with different libraries used to do the threading, i.e. OpenMP or POSIX Threads.
	Nevertheless, splitting into host and device enforces cleaner and more performant code, which we see also in our benchmarks.
\end{itemize}
As the actual hardware/software pair used for computations is abstracted away, a small set of common operations is sufficient to dispatch computations to any kind of device.

To iterate over a full lattice, we provide two functions for common operations, shown below:
\begin{itemize}
	\item \cpp{TempLat::device::iteration::foreach} takes a functor, e.g. a lambda function, and dispatches it over the whole lattice. This can be used for memory transformations --- the assignment operators of \cpp{Field} and other classes dispatch the evaluation of expressions in precisely this way, as shown in \cref{lst:foreach}. In this example from \bash{lattice/field/views/fieldviewconfig.h}, one sees that the dispatching function \cpp{TempLat::device::iteration::foreach} additionally takes a name that uniquely identifies the operation (for debugging purposes) and a layout which specifies the lattice. \cpp{DoEval::eval} evaluates the expression \cpp{g} at the \cpp{NDim}-dimensional index \cpp{idx}.
	\item \cpp{TempLat::device::iteration::reduce} similarly takes a functor and performs a reduction operation. As an example, we show a maximum over a lattice in \cref{lst:reduce}, which is taken from \bash{lattice/measuringtools/maximum.h}.
\end{itemize}
Besides pure iteration, memory management operations are also represented by \TempLat:
\begin{itemize}
	\item An actual (owning) block of memory is represented by \cpp{TempLat::device::memory::NDView}. This class owns the memory and provides multi-index access, similar to the C++23 \cpp{std::mdspan}. It can also mirror the memory to the host on-demand.
	\item \cpp{TempLat::device::memory::copyDeviceToDevice} performs a memory copy on the device itself.
	\item \cpp{TempLat::device::memory::copyHostToDevice} performs a   memory copy from host to device.
	\item \cpp{TempLat::device::memory::copyDeviceToHost} performs a   memory copy from device to host.
\end{itemize}
All operations work asynchronously with the host code. As a consequence, synchronization has to be enforced explicitly --- otherwise the device will perform all queued operations in order, but independently of host work.\\
To synchronize host and device, we provide the function \cpp{TempLat::device::iteration::fence}.

\subsection{Iterating the lattice on different hardware}
\label{sec:iteration}

\begin{figure*}[t]
	\centering
	\begin{minipage}[]{0.4\linewidth}
		\centering
		\includegraphics[width=0.8\linewidth]{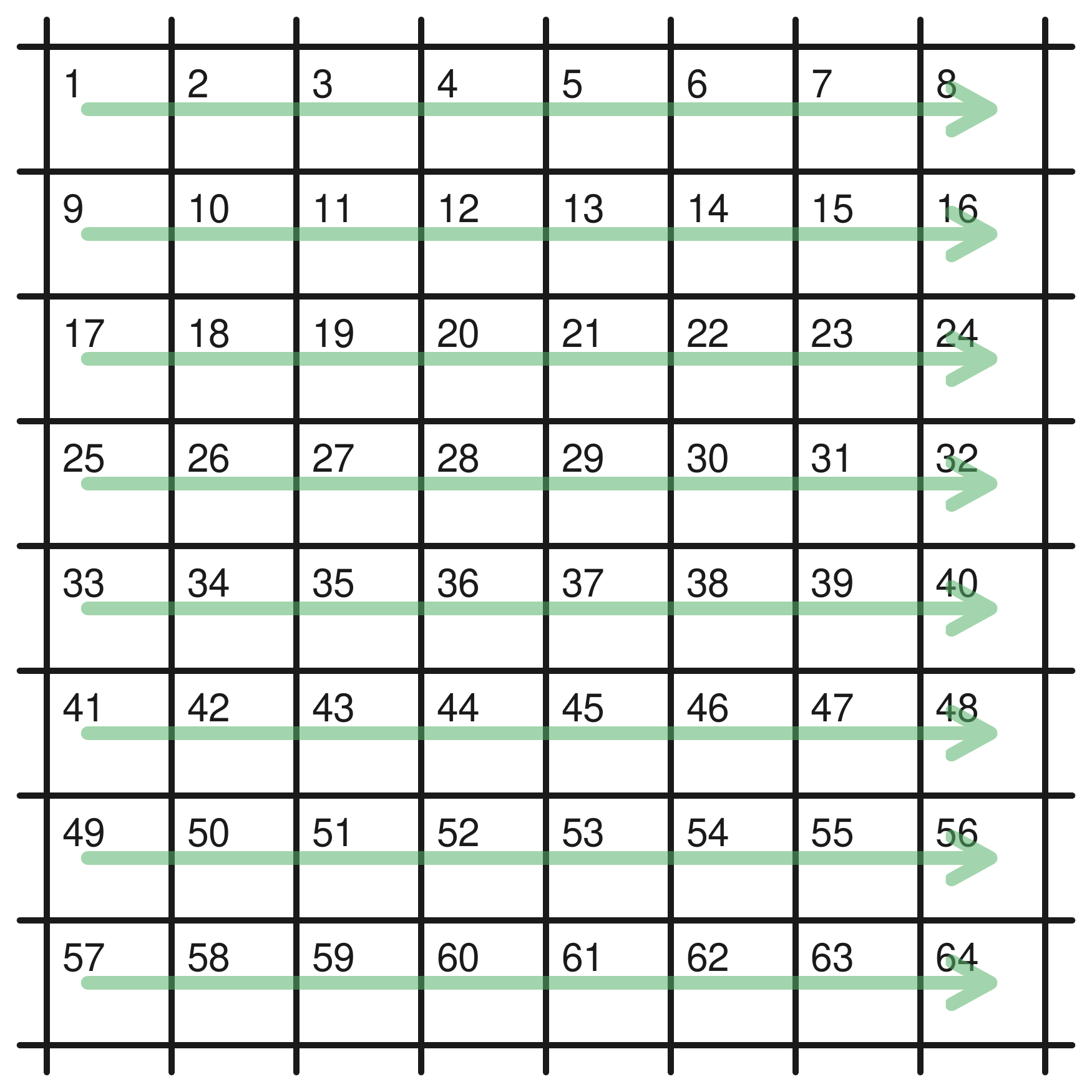}
	\end{minipage}%
	\begin{minipage}[]{0.4\linewidth}
		\centering
		\includegraphics[width=0.8\linewidth]{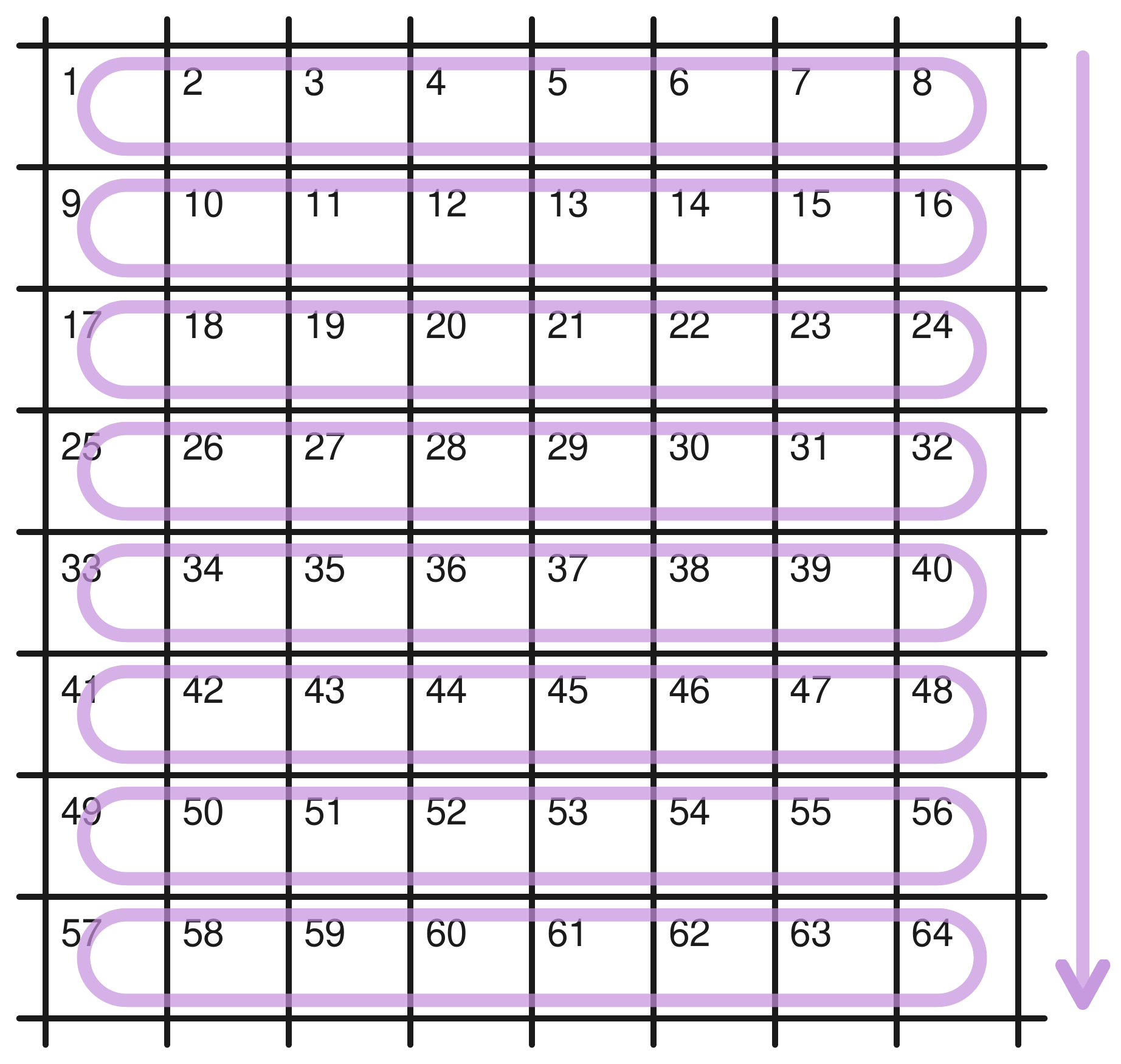}
	\end{minipage}
	\caption{\textbf{Left}: Cached memory access: each thread gets a block of memory it walks through, allowing it to both cache the memory and enable SIMD. \textbf{Right}: Coalesced memory access, where a warp of threads walks through the memory. Each thread in the warp thus has a fixed stride when iterating. For illustration we show warp sizes of 8, but NVIDIA hardware typically has warp sizes of 32, while AMD wavefronts are 32 or 64 wide.}
	\label{fig:caching_coalescing}
\end{figure*}

When writing performance-portable code for both CPU and GPU it is important to note that the optimal memory access patterns, i.e. those that maximize throughput, differ between the hardware architectures.

Specifically, on a CPU the optimal way to traverse memory is one that allows for caching within a single thread. So, a single thread should get a contiguous chunk of memory to work through, which allows both for caching and for SIMD vectorization optimizations to be used. This is the \textit{cached} memory access pattern shown in the left side of \cref{fig:caching_coalescing}.

On a GPU, the paradigm is different: multiple threads in the smallest thread group structure on a GPU, a so-called warp, always execute the same operation in tandem. Effectively, what optimally happens in the warp is that source memory is read into local cache as a warp-sized block, and then each thread in the warp operates on one memory location within the cached block. Then, the warp proceeds to the next block and repeats this. 
The memory access pattern is thus \textit{coalesced} and each thread jumps a certain fixed stride length in memory as it continues to work.
This memory access pattern is shown in the right side of \cref{fig:caching_coalescing}.

In contrast to how \Kokkos implements this hardware-dependent iteration, we keep the memory layout row-major on all hardware, and instead change only the iteration order: row-major (cached) on CPU, column-major (coalesced) on GPU. This choice is made at compile time, depending on which hardware has been configured for the library.

\subsection{Expression templates}
\label{sec:expressiontemplates}

We have already introduced the concept of expression templates in \cref{sec:interface}. Beyond the convenient and versatile interface this offers, the design of having an expression that is then dispatched onto the device allows for performance optimizations that would be otherwise impossible.

Due to the full expression tree being available at compile time, an evaluation, usually invoked through an assignment operation, fuses the entire expression tree into a single kernel. This is in contrast to what would happen if the algebra were actually implemented as operations on memory: every operation would be performed in parallel on the lattice, but in separate sequential steps.
Fusing the expression tree allows the compiler to remove temporaries, perform common subexpression elimination within the tree and avoid allocations. As such, even the $SU(2)$ algebra is performed only with temporaries on the stack.

Furthermore, the expression tree allows us to automatically dispatch DFTs and halo updates (ghost exchanges), depending on the type of expression. The only operators needing ghosts are shifts or other derivative operators. On evaluation of an expression tree, \TempLat recurses the full tree at compile time and decides whether the evaluation requires an update to the halo.
The same applies to Fourier transformations: whenever an assignment or an expression requires the field to be in its Fourier representation, a DFT is automatically performed before the assignment itself.

Operators such as Levi-Civita symbols or Kronecker deltas are easily expressed as fully compile-time arbitrary-dimension objects with the help of \cpp{ZeroType} and \cpp{OneType}, making the direct usage of these feasible without needing to think about sparsity of expressions.
Other numeric operators are also tightly optimized: for example, integer powers are computed using the fewest machine instructions possible.

We emphasize again that this machinery is also easy to extend, as each group structure (\cpp{Field}, \cpp{Matrix}, \cpp{SU2}, \cpp{SU2Doublet}, \cpp{ComplexField}) defines its own algebra. The operators inspect the structure of their operands: \cpp{operator*}, for instance, resolves to different functions depending on whether one multiplies two $SU(2)$ fields, two complex fields, or an $SU(2)$ field with a complex field. This paves the way for future extensions to $SU(3)$ or other groups.

\subsection{Fourier Transformations}
\label{sec:fourier}

Fourier transformations are an essential tool both for state preparation and measurements in lattice simulations. As explained in \cref{sec:interface}, \TempLat is envisioned as both a dimension-agnostic and performance-portable lattice simulation toolkit. This entails the need for a library providing discrete Fourier transforms (DFTs) in arbitrary dimensions that works on all supported hardware (CPUs and AMD/NVIDIA GPUs) with hybrid parallelization.

While some libraries that support distributed DFTs on GPUs exist, e.g. \texttt{heFFTe}~\cite{heffte}, cuFFTMp (part of the NVIDIA HPC SDK), and several others, these either support only one hardware vendor or are limited to 2 or 3 dimensions. 

As there is currently no library available that has both arbitrary-dimension support and performance portability with distributed support, we developed the DFT library \ParaFaFT based on the algorithm presented in \cite{DBLP:journals/corr/abs-1804-09536}. \ParaFaFT is available on GitHub\footnote{\url{https://github.com/cosmolattice/parafaft}} and supports any dimensionality $d\geq2$. 
\ParaFaFT handles the domain decomposition and distribution algorithm, while dispatching local DFTs to different backends. Depending on the hardware used, the library supports three backends: FFTW3~\cite{FFTW05} on CPU, cuFFT~\cite{cufft} on NVIDIA GPUs, and hipFFT~\cite{hipfft}, which forwards to rocFFT, on AMD GPUs.

\begin{figure*}[pos=t]
	\centering
	
	\includegraphics[valign=t]{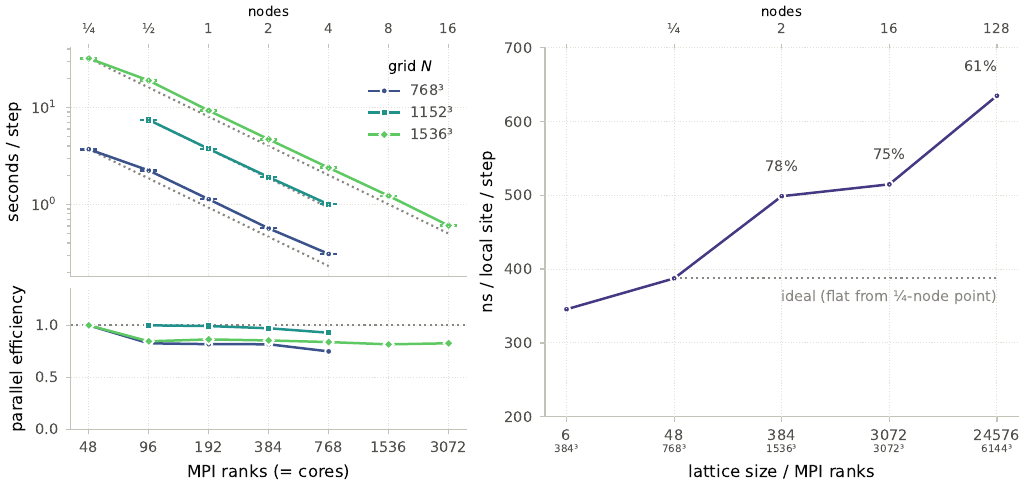}%
	\caption{Strong (left) and weak (right) scaling behavior of \TempLat on the CPU partition of the Otus cluster.}
	\label{fig:templat_mpi_scaling}
	
	\vspace{10pt}
	
	\includegraphics[valign=t]{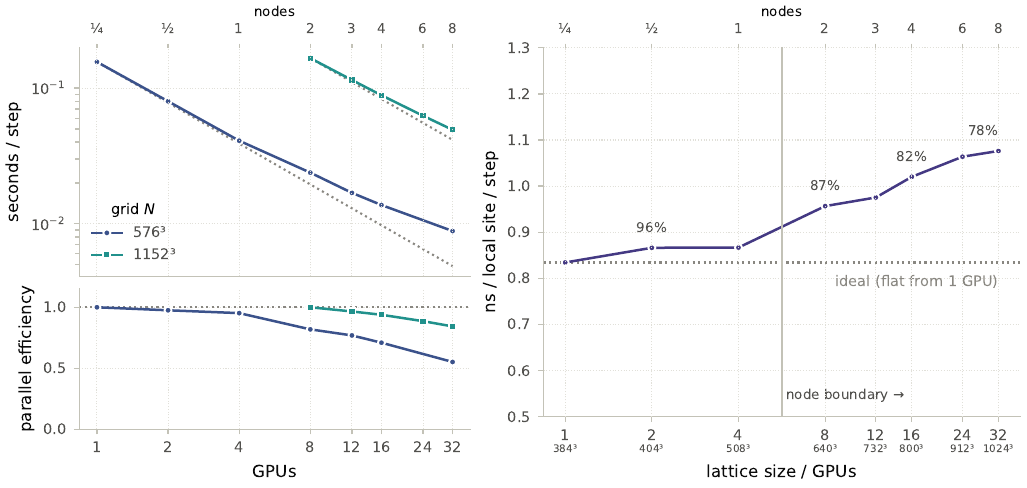}%
	\caption{Strong (left) and weak (right) scaling behavior of \TempLat on the A100 GPUs of the Noctua2 cluster.}
	\label{fig:templat_gpu_scaling}
\end{figure*}

\section{Benchmarks}
\label{sec:bench}


We evaluate the performance and scaling behavior of \TempLat on the Otus HPC cluster \cite{otus2025} (CPU) and the Noctua2 HPC cluster \cite{noctua2} (GPU) of the Paderborn Center for Parallel Computing (PC2). A CPU node of Otus consists of two AMD EPYC 9655 CPU sockets, with 96 Zen5 cores each.
A GPU node on Noctua2 has four NVIDIA A100 40 GB GPUs, connected via NVLink.

As a benchmark workload, we evolve a classical $SU(2)$ Yang--Mills field in $d=3$ with a leapfrog integrator in temporal gauge ($A_0=0$), evolving the gauge links $U_i^a$ and the electric fields $E_i^a$ in tandem, see for instance \cite{Figueroa:2020rrl}.
We use double precision and periodic boundary conditions with a single ghost layer. 
This exercises every part of \TempLat, from the expression algebra itself to the ghost updates and distribution strategy. Each time step consists of three phases:
\begin{itemize}
	\item \textbf{halo}: exchange the link ghost layers across all directions (pure MPI/device communication);
	\item \textbf{kick}: $E_i^+ = E_i - \mathrm{d}t\,\sum_j(P_{ij} - P_{ij}^\dagger)$ ($SU(2)$ products and algebra projection, stencil-heavy);
	\item \textbf{drift}: $U_i^+ = \exp(\mathrm{d}t\, E_i)\,U_i$ (closed-form $SU(2)$ exponential map and group multiply).
\end{itemize}

\noindent In the above, $E_i^+$ and $U_i^+$ are the electric fields and links at the next time step, respectively. After 10 untimed warm-up steps, we accumulate the timings over 100 timed steps.
We report the wall-clock time per lattice site per step (in nanoseconds), which is directly comparable across lattice sizes. We also measure each of the phases separately. On the GPU side, we place one MPI rank per A100, with NVLink for intra-node and InfiniBand for inter-node communication.

We perform both a strong- and a weak-scaling analysis. For the strong-scaling analysis we keep the global lattice fixed and increase the number of ranks; for the weak-scaling analysis we keep the local volume per rank fixed and grow the global lattice with the rank count. The configurations are summarized below:
\begin{table}[pos=H]
	\centering
	\begin{tabular}{@{} p{2.2cm} p{2.8cm} p{2.6cm} @{}}
		\toprule
		 & \textbf{Strong} & \textbf{Weak} \\
		\midrule
		CPU (Otus) & $N=768,\,1152,\,1536$ & $9.4\times10^{6}$~sites/core\\
		\midrule
		GPU~(Noctua2) & $N=576,\,1152$ & $3.3\times10^{7}$ sites/GPU\\
		\bottomrule
	\end{tabular}
\end{table}

\noindent
The resulting timings are shown in \cref{fig:templat_mpi_scaling} for the CPU and \cref{fig:templat_gpu_scaling} for the GPU.

On the CPU side, the strong scaling is almost perfectly linear as soon as the memory bandwidth of a single node is saturated. The halo-exchange penalty becomes noticeable only for a very large number of processes, when each process becomes computationally starved.

This effect is more noticeable in the GPU data for the smaller lattice. While the intra-node scaling is almost perfect thanks to the NVLink connection, the performance slowly degrades as traffic goes through the InfiniBand, until each GPU no longer holds enough of the lattice to saturate its own memory bandwidth.
The weak-scaling analyses tell a similar story, with performance degrading as more and more data is exchanged over the InfiniBand. Unsurprisingly, the effect is similar across CPU and GPU, as it comes from the MPI communications. Within a single GPU node, the NVLink keeps the intra-node communication cost negligible even as the global lattice grows.

\begin{figure}[pos=t]
	\centering
	\includegraphics{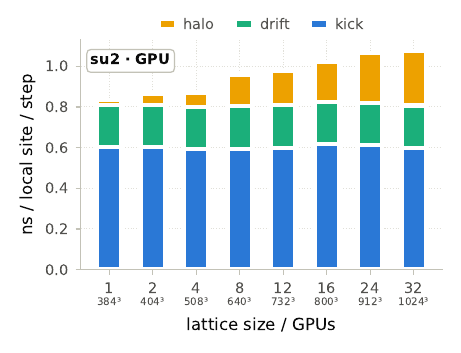}%
	\capfitwidth
	\caption{Decomposition of a single step in the $SU(2)$ benchmarks of the weak-scaling plot of \cref{fig:templat_gpu_scaling}. Kick and drift show optimal behavior --- their cost per site stays constant. The halo cost, in contrast, grows with the number of GPUs, since dividing the lattice further increases the surface-to-volume ratio of each local domain.}%
	\label{fig:su2timingphases}%
\end{figure}

We further analyze this in \cref{fig:su2timingphases}, where we decompose the time spent in each phase. The kick and drift behave optimally: their timings stay constant, no matter how many GPUs we use. The halo cost, on the other hand, increases. This can be heuristically understood as follows.
\TempLat always inherits the domain decomposition used by the accompanying Fourier transformation library --- in this case, \ParaFaFT, which uses a pencil decomposition: one dimension is kept intact and the remaining $d-1$ are segmented, yielding long ``pencils''.
In $d=3$, an ideal partition into $P$ processes on a $\sqrt{P}\times\sqrt{P}$ grid requires $4N^2P^{-1/2}$ sites to be communicated per process, assuming a single ghost layer and neglecting the top and bottom of the pencils. Since the weak-scaling analysis keeps $N^3 / P$ fixed, the total communication per lattice site scales at least as $\mathcal{O}(P^{1/6})$.
The data shown in \cref{fig:su2timingphases} exhibit a scaling behavior closer to $\mathcal{O}(P^{1/2})$; the difference is due to the use of non-square decompositions and additional synchronization overhead.

We end this section by leveraging one of the strengths we mentioned in the introduction: the \HILA code can be used to perform the same computations, allowing us to validate the scaling of \TempLat across nodes. For this comparison, we switch to the simplest kernel that still exercises the same machinery and can be reproduced unambiguously in two independent codebases: a single real scalar field with quartic self-interaction in $d=3$, evolved with the same kick--drift leapfrog, again in double precision with periodic boundary conditions and a single ghost layer. Each step consists of
\begin{itemize}
	\item \textbf{halo}: exchange the ghost layers of $\phi$ (pure MPI/device communication);
	\item \textbf{kick}: $\pi^+ = \pi + \mathrm{d}t\,(\Delta_{\rm lat}\phi - m^2\phi - \lambda\phi^3)$ (7-point stencil and pointwise arithmetic);
	\item \textbf{drift}: $\phi^+ = \phi + \mathrm{d}t\,\pi^+$ (pointwise, bandwidth-bound).
\end{itemize}

\noindent We start from Gaussian white-noise initial conditions. Each phase is written in the native idiom of the respective code, and the timing protocol is the same as before: 10 untimed warm-up steps followed by 100 timed steps, with every phase fenced by a device synchronization and an MPI barrier, and one MPI rank per A100. The strong-scaling analysis uses global lattices of $768^3$ and $1536^3$, while the weak-scaling one keeps $\approx 512^3$ sites per GPU, from $512^3$ on a single GPU up to $1632^3$ on 32.

\begin{figure*}[pos=t]
	\centering
	\includegraphics{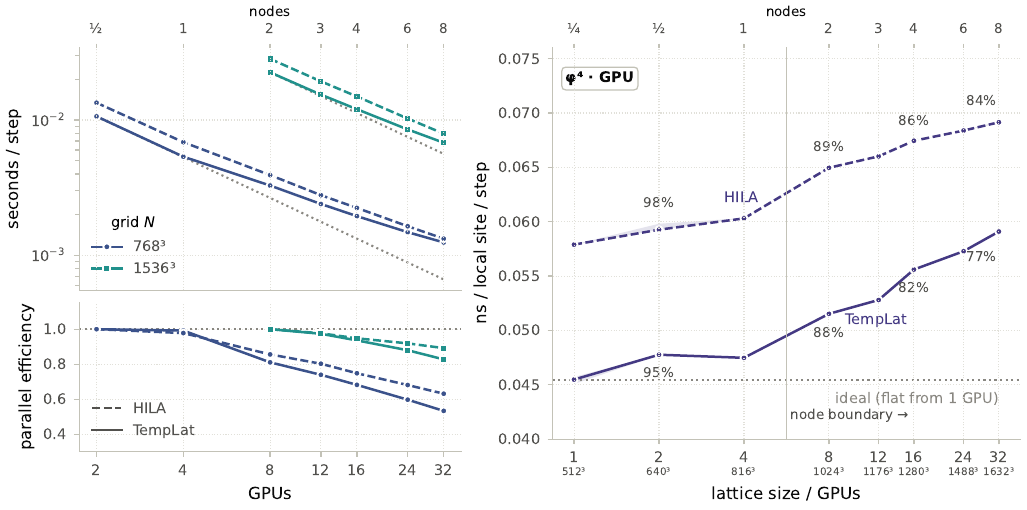}%
	\capfitwidth
	\caption{Strong (left) and weak (right) scaling behavior of \TempLat and \HILA on the A100 GPUs of the Noctua2 cluster. For this scaling analysis we use a classical relativistic scalar theory, evolved with a leapfrog algorithm. While we see very similar scaling curves for \TempLat and \HILA, in strong scaling we find that at high GPU counts, the block decomposition of \HILA seems to scale better than the pencil decomposition of \TempLat.}%
	\label{fig:hila}%
\end{figure*}

The weak-scaling analysis, shown on the right-hand side of \cref{fig:hila}, demonstrates that the two completely independent codes scale in a very similar way, with a moderate offset in \TempLat's favor. \HILA, however, scales better at large node numbers. This is most apparent in the strong-scaling analysis on the left-hand side, where we push to a large number of GPUs and each GPU becomes correspondingly underloaded: we observe a crossing, with \TempLat starting again from a constant offset in its favor and \HILA eventually compensating through its better scaling. The improved scaling behavior can be traced to more aggressive optimization of the MPI exchanges and to a different domain decomposition: \HILA decomposes in $d$ dimensions instead of $d-1$. Keeping $N$ fixed, we expect a total communication cost $\sim 4 N^2 P^{1/2}$ for a pencil decomposition, whereas for a block decomposition we expect $\sim 6 N^2 P^{1/3}$, again assuming decompositions into grids of shape $\sqrt{P}\times\sqrt{P}$ or $P^{1/3}\times P^{1/3}\times P^{1/3}$. Note that the decomposition strategy does not explain the constant offset in \TempLat's favor, as it is already present with a single GPU; that offset remains to be fully explained and reproduced across hardware and compiler stacks. This gives us a clear direction to further optimize \TempLat at scale, and illustrates how future improvements of both codes will be partially driven by more thorough comparisons.

\section{Conclusions}
\label{sec:conclusions}

We motivated the need for a versatile, user-friendly, and highly efficient lattice field theory framework, and showed that \TempLat delivers on all of these fronts.
We presented how its expression-template-based algebra enables transparent and quick implementations of a variety of physical systems and algorithms in arbitrary dimensions; this feature remains at the heart of \CosmoLattice's success.
The algebra fully decouples the symbolic content of an expression from its evaluation; we showed how we leverage this separation, via \Kokkos, to generate highly performant code for a wide range of hardware. We substantiated these claims with benchmarks on both CPUs and GPUs, demonstrating excellent scaling of the code to many nodes.
Finally, we illustrated the complementarity with \HILA, as advertised in the introduction, comparing the performance of both codes on a simple $\phi^4$ theory, and explaining how such cross-validation and stress-testing will improve both of them.

We also benchmarked our standalone parallel DFT library \ParaFaFT, which parallelizes multidimensional DFTs with an efficient pencil decomposition.
While \ParaFaFT performs well on CPUs and is rarely the bottleneck in \TempLat applications, its GPU performance falls behind \texttt{cuFFTMp} at smaller lattice sizes.
We traced this to our lattice decomposition being restricted to pencils; decoupling the real-space and Fourier-transform geometries is a promising avenue, already exploited successfully by \HILA. This would allow us to pick, depending on lattice and MPI sizes, optimal decompositions separately in Fourier- and position-space.
We also plan to optimize our MPI communications more aggressively, with non-blocking exchanges that overlap bulk and halo work.
Thanks to the abstraction of the device model used in \TempLat, we also envision broader hardware support in the future, e.g.\ through SYCL (Intel GPUs) and Metal (Apple SoCs).
Beyond these technical improvements, we plan to further develop the physics algebra, adding dedicated $SU(N)$ algebras and support for simulating semiclassical fermions \cite{Aarts:1998td}.
A Monte Carlo lattice framework driven by \TempLat is also under development.

\section*{Acknowledgments}

The authors thank the CosmoLattice team (J.~Baeza-Ballesteros, D.~G.~Figueroa, N.~Loayza, F.~Torrenti, A.~Urio, W.~Valkenburg) for many discussions and tests of the software. They are particularly grateful to J.~Baeza-Ballesteros for his direct contributions to the matrix algebra of \TempLat, and to W.~Valkenburg, who crafted the very first draft from which \TempLat eventually emerged. They are also grateful to K.~Rummukainen for comments on a version of this manuscript. A.F. thanks S.~Schlichting for encouraging him to start this project and J.~Constantinides, E.~Grossi and L.~Mazur for numerous discussions on HPC. The authors are funded
by the Deutsche Forschungsgemeinschaft (DFG, German Research Foundation) through the Emmy
Noether Programme Project No. 545261797.
The authors gratefully acknowledge the computing time made available to them on the high-performance computers Otus and Noctua2 at the NHR Center Paderborn Center for Parallel Computing (PC2). This center is jointly supported by the Federal Ministry of Research, Technology and Space and the state governments participating in the National High-Performance Computing (NHR) joint funding program (\url{www.nhr-verein.de/en/our-partners}).

\appendix

\section{ParaFaFT}
\label{sec:app_parafaft}

\begin{figure*}[pos=t]
	\centering 
	
	\includegraphics{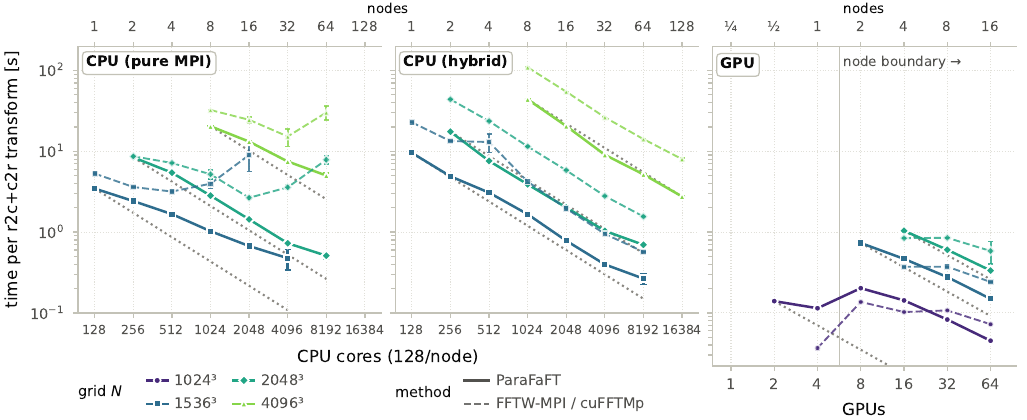}%
	\caption{Strong scaling behavior of \ParaFaFT on the Noctua2 cluster, for three-dimensional \texttt{r2c}+\texttt{c2r} transforms. Left: CPU runs with pure MPI parallelization. Middle: CPU runs with hybrid MPI+OpenMP parallelization. Right: GPU runs on A100s. In each panel \ParaFaFT (solid) is compared against the corresponding baseline (dashed), i.e.\ MPI-parallel FFTW3 on CPU and \texttt{cuFFTMp} on GPU. Dotted lines indicate ideal scaling.}
	\label{fig:parafaft_strong_scaling}
\end{figure*}

We benchmark \ParaFaFT on the Noctua2 HPC cluster~\cite{noctua2} of the Paderborn Center for Parallel Computing. Each CPU node has two AMD EPYC 7763 sockets with 64 Zen3 cores each; each GPU node has four NVIDIA A100 40 GB GPUs connected via NVLink.

We perform the DFTs in the following configurations:
\begin{table}[pos=H]
	\centering
	\begin{tabular}{@{} C{3.5cm} C{3.5cm} @{}}
		\toprule
		\textbf{Configuration} & \textbf{Lattice sizes} \\
		\midrule
		\ParaFaFT sequential & $N=1536,\,2048,\,4096$ \\
		\ParaFaFT with OpenMP & $N=1536,\,2048,\,4096$ \\
		\ParaFaFT with GPU & $N=1024,\,1536,\,2048$ \\
		\midrule
		\texttt{FFTW3} sequential & $N=1536,\,2048,\,4096$ \\
		\texttt{FFTW3} with OpenMP & $N=1536,\,2048,\,4096$ \\
		\texttt{cuFFTMp} & $N=1024,\,1536,\,2048$ \\
		\bottomrule
	\end{tabular}
	\label{tab:parafaft_configurations}
\end{table}

\noindent
We perform three-dimensional real-to-complex (forward) and complex-to-real (backward) DFTs on lattices with sizes $N=1024,\,1536,\,2048,\,4096$.
After a single warm-up cycle, each data point averages 50 forward--backward transform cycles; we quote the mean and standard deviation.
The strong-scaling analysis fully occupies 1 to 128 nodes, i.e.\ 128 to 16384 CPU cores; for the GPU comparison we equate one CPU node with a single GPU. The resulting timings are shown in \cref{fig:parafaft_strong_scaling}.

In the left two plots of \cref{fig:parafaft_strong_scaling} we see that a hybrid parallelization strategy leads to almost perfect scaling for purely CPU runs, whereas the MPI implementation shows a flatter scaling profile.
In comparison to the MPI parallelization of FFTW3, the MPI-variant of \ParaFaFT scales much better. This is due to its usage of pencil decompositions, which are more optimal than slab decompositions at very high MPI rank counts. In all configurations \ParaFaFT also exhibits a fixed offset performance improvement with respect to FFTW3.

The right panel of \cref{fig:parafaft_strong_scaling} shows the GPU scaling behavior. \ParaFaFT is slower than the \texttt{cuFFTMp} baseline for smaller lattices, while it overtakes it at increasing numbers of GPUs.
The reason here is opposite to the CPU case: \ParaFaFT again defaults to pencil decompositions, which are more optimal for ghosting, but require more transposition work for a multidimensional DFT.
The \texttt{cuFFTMp} runs use slabs, which keep two dimensions local and split only the third, and are therefore more efficient for DFTs at smaller GPU counts.

Where DFTs are performance-critical, this points to a future improvement of the geometry handling: performing them in a slab decomposition and redistributing to pencils for all other purposes.


\bibliographystyle{elsarticle-num}

\bibliography{ref-lib}



\end{document}